\documentstyle[12pt,aaspp4]{article}
\def\arcsec{^{\prime \prime}}
\def\arcmin{^{\prime}}
\def\deg{^{\circ}}
\def\sun{\odot}

\def\be{\begin{equation}}
\def\ee{\end{equation}}
\def\etal{et~al.}
\def\kms{km\,s$^{-1}$}
\def\vsys{$v_{\rm sys}$}

\def\eg{{\it e.g.}}
\def\Z35{IIIZw35}
\def\msun{{\rm M}_\sun}
\def\lsim{\mathrel{\raise .4ex\hbox{\rlap{$<$}\lower 1.2ex\hbox{$\sim$}}}}
\def\gsim{\mathrel{\raise .4ex\hbox{\rlap{$>$}\lower 1.2ex\hbox{$\sim$}}}}
\def\inv{\ifmmode^{-1}\else$^{-1}$\fi}

\begin{document}
\title{VLBA Imaging of the OH Maser in IIIZw35}

\author{Adam S. Trotter, James M. Moran, \& Lincoln J. Greenhill}
\affil{Harvard--Smithsonian Center for Astrophysics, MS 42, 60 Garden Street,
Cambridge, MA 02138}
\author{Xing--Wu Zheng}
\affil{Department of Astronomy, Nanjing University, 
Nanjing, Jiangsu, China 210093}
\author{Carl R. Gwinn}
\affil{Department of Physics, University of California,
Santa Barbara, CA 93106}
\vspace{2in}
\begin{center}
Submitted to {\it The Astrophysical Journal Letters}, 14 April 1997

Accepted 6 June 1997
\end{center}

\begin{abstract}
We present a parsec--scale image of the OH maser in the nucleus of 
the active galaxy IIIZw35, made using the Very Long Baseline Array
at a wavelength of 18\,cm.
We detected two distinct components, with a projected separation of
50\,pc (for $D=110$\,Mpc) 
and a separation in Doppler velocity of 70\,\kms, which contain
50\% of the total maser flux.
Velocity gradients within these components 
could indicate rotation of clouds with binding mass densities
of $\sim 7000\,\msun$\,pc$^{-3}$ and total masses
$\gsim 5\times 10^5\,\msun$.  Emission in the 1665--MHz OH
line is roughly coincident in position
with that in the 1667--MHz line, although
the lines peak at different Doppler velocities.  
We detected no $\lambda =18$\,cm continuum emission; 
our upper limit implies a peak apparent optical depth
$|\tau_{\rm OH}| > 3.4$, assuming the maser is an unsaturated
amplifier of continuum radiation.

\end{abstract}

\section{Introduction} \label{intro}
\Z35 is a pair of galaxies catalogued in Zwicky (1971)
as a 16th--magnitude compact blue double.  
Its systemic velocity, 
based on observations of CO emission (\cite{mir87}),
is 8289\,\kms\ (heliocentric, optical definition),
corresponding to a distance of 110\,Mpc ($H_0 =75$\,\kms\,Mpc\inv;
1\,mas is equivalent to a projected separation of 0.53\,pc).
Optical and IR images 
(\cite{cha90}) show two components separated by $9\arcsec$
along a position angle of $20\deg$.  In optical
images, the northern
galaxy is brighter and redder than the southern,
and emits the majority of its energy in the far infrared;
{\it IRAS} measured its $10-120\,\mu$m luminosity to be 
$3.5\times 10^{11}\,{\rm L}_{\sun}$, with a spectral energy distribution
consistent with thermal
emission from dust at 65\,K.  Chapman \etal\ (1990) classify the northern
galaxy as either a LINER or
Seyfert~2, based on optical emission line ratios.

The nucleus of the
northern component of \Z35 is the site of strong OH maser emission.
The total luminosity in the 1667--MHz transition, assuming isotropic
emission, is $530\,{\rm L}_\sun$, with
a ratio of flux density in the 1667 and 1665--MHz spectral peaks of
9:1 (\cite{sta87}).  
The 1667--MHz spectrum
has two peaks at approximately $\pm 35$\,\kms\ from the systemic velocity.
The maser extends in Doppler velocity over about 100\,\kms, but there is also
faint ($< 10$\,mJy) blueshifted emission extending more than 500\,\kms\ from
\vsys.  Baan, Haschick, \& Henkel (1989) interpret the
blueshifted wings in this and other OH maser galaxies
to be due to molecular outflows.  

Chapman \etal\ (1990) imaged the
$\lambda=18$\,cm continuum emission from \Z35 with
the VLA and MERLIN and measured a flux density of
$30\pm 10$\,mJy and an angular size of $\lsim 300$\,mas ($T_B > 10^3$\,K).
The centroid of the maser emission 
was observed to coincide with that of the 18\,cm continuum
to within $0.3\arcsec$.
Assuming the maser acts as an unsaturated
amplifier and completely covers the continuum,
they inferred an apparent peak OH
optical depth $|\tau_{\rm OH}|\sim 1.7$.   
Montgomery \& Cohen (1992) mapped the OH maser emission
with MERLIN and observed an extended ($\sim 80$\,mas) structure
with an apparent velocity
gradient of 2.5\,\kms\,pc\inv\ along a position angle of $-20\deg$.
They interpreted the gradient to be the signature of
a rotating molecular disk in the nucleus of \Z35.   

\section{Observations} \label{obs}

We observed \Z35 on 25 March 1995 at $\lambda=18$\,cm with the 
NRAO Very Long Baseline Array (VLBA)\footnote{The National 
Radio Astronomy Observatory
is a facility of the National Science Foundation operated under
cooperative agreement with Associated Universities, Inc.}
in spectral--line mode.
Only the stations at Fort Davis, Kitt Peak, Los Alamos, North Liberty,
Owens Valley and Pie Town provided useful data.
The baseline fringe spacings ranged from 17 to 200\,mas.
The data were recorded in both left and right circular polarization, in
two 8--MHz passbands centered on $v = 8438$ and 7000\,\kms.  The
former covers both the 1667--MHz and the 1665--MHz transitions of
OH, while the latter
was tuned to lie outside the known
range of maser emission.  The data were
correlated at NRAO; each passband was divided into 512
spectral channels, yielding a Doppler velocity spacing of 2.8\,\kms.  
All further processing was undertaken with
AIPS at the CfA.  After initial calibration of the amplitudes, fringe rates,
and delays, the visibilities were phase--referenced to the channel
of the peak line emission, at 8245\,\kms.  
Analysis of residual fringe rates established the position of the
maser peak: 
$\alpha_{1950}=01^{\rm h}41^{\rm m}47^{\rm s}\!\!.925\pm 0^{\rm s}\!\!.007$,
$\delta_{1950}=16\deg 51\arcmin 05\arcsec\!\!.8\pm 0\arcsec\!\!.2$.
An error in this reference position during processing resulted in
a systematic error in the positions of spectral features that scales
linearly with frequency offset from the maser peak.  The magnitude
of this error is $1.7$\,mas\,MHz\inv; the resulting systematic errors are
$<1$\,mas in the 1667--MHz line profile,
and $<4$\,mas in the 1665--MHz line profile.
The 1--$\sigma$ noise level in a 
uniformly--weighted spectral channel image was 3.8\,mJy/beam.
Spectral channels in the lower--frequency passband were averaged
to produce continuum visibilities.
The 1--$\sigma$ noise level in a uniformly--weighted
continuum image was 0.18\,mJy/beam.  

\section{Results} \label{res}

Figure~1 shows the distribution of 1667--MHz
OH maser emission in the nucleus of \Z35. 
We detected emission in two clumps, with a projected
separation of 90\,mas (50\,pc)
along an axis with a position angle of $-10\deg$.  
The spectral peak of the maser emission is in the southern clump.
Figure~2 shows expanded views of the northern and southern clumps.
The positions of the 1665--MHz line peaks are also plotted.
In the single--channel images, the maser features were unresolved;
we place an upper limit of 17\,mas on their angular sizes, which
implies a peak brightness temperature $T_{\rm OH} > 3\times 10^6$\,K.  
Figure~3 compares the imaged flux density in both
clumps with the single--dish
spectrum of Staveley--Smith \etal\ (1987).  
Note that $\sim 50\%$ of
the total maser flux density was resolved in our VLBA images.  
Sensitivity limitations did not permit us to
image the full range of blueshifted emission, which
extends $\sim 500$\,\kms\ from the systemic velocity 
at a level of less than 10\,mJy (\cite{baa89}).

The northern clump has a projected extent of $\sim 10$\,pc (Figure~2a).  
It consists of multiple spectral features,
over a velocity range of $110$\,\kms\ (Figure~3).
We observed an east--to--west velocity gradient of 11\,\kms\,pc\inv\
over $\sim 5$\,pc in the velocity range $8290<v<8340$\,\kms\ 
(Figure~1, {\it inset}).
Between 8230 and 8290\,\kms, we did not detect a gradient; 
all of the emission in this range
lies in a compact ($\lsim 2$\,pc) cluster at the
SE end of the clump (Figure~2a).
A recent high--resolution intercontinental VLBI observation
revealed a north--south
gradient in this cluster of $30$\,\kms\,pc\inv\ (P. Diamond, private
communication). 
Lower sensitivity and angular resolution prevented us from detecting
a gradient of this magnitude.

The brightest emission is in the southern clump.
The imaged line profile is well fit by a Gaussian function
with a peak of 130\,mJy at 8245\,\kms\ and a FWHM of 30\,\kms, 
with line wings extending $\pm 50$\,\kms\ from the peak at a 
level of $<50$\,mJy. 
Most of the emission in the southern
clump is confined to a region approximately 3\,pc across (Figure~2b);
however, emission in the line wings
extends $\sim 5$\,pc SE and NW of the maser peak. 
Evidence for a velocity gradient is ambiguous within the compact
part of the southern clump, but emission from the line wings
suggests
a velocity gradient of about 10\,\kms\,pc\inv, increasing
along a position angle of $110\deg$. This gradient is of roughly
the same magnitude as that in the northern clump, but in
the opposite direction (Figure 1).

We imaged 1665--MHz maser emission
in both the northern and southern clumps.
The southern clump
has two spectral peaks, at 8232 and 8258\,\kms, both with 
flux densities of $\sim 10$\,mJy.
These two peaks straddle the 1667--MHz peak in Doppler velocity (Figure~3).
Both of the southern 1665--MHz features are coincident with the 
compact component of the 1667--MHz
clump within the position errors (Figure~2b).  In the northern clump,
we detected one
1665--MHz feature, also with a peak flux density of $\sim 10$\,mJy.  The
velocity of this feature is 8353\,\kms, which is near the red edge of
the 1667--MHz profile.
Moreover, while the velocity of the 1667--MHz emission increases to
the west in the northern clump, this 1665--MHz feature
lies at its eastern end (Figure~2a).  It appears that the
two OH maser clumps differ in their relative distributions of
1667 and 1665--MHz emission, both in angle and in velocity.
Note that the features we attribute to the 1665--MHz
transition could be redshifted ($\sim 8600$\,\kms)
1667--MHz features.  
Single--dish spectra (\eg, \cite{mir87})
do show faint emission filling the gap between
the main 1667--MHz profile and the putative 1665--MHz profile.
However, the near angular and Doppler velocity coincidence 
of emission in the two profiles
suggests we are detecting
the 1665--MHz transition in both clumps, rather than highly 
redshifted 1667--MHz features.

We did not detect any 18\,cm continuum
emission in the nucleus of \Z35 with the VLBA.
We imaged various subsets of the data with a variety of
weighting schemes 
to confirm this non--detection.  We place a
$5\sigma$ upper limit on the continuum flux density 
of 1\,mJy/beam.
The visibility amplitudes, averaged over the entire observation, were
consistent with $\sigma =1$\,mJy noise on all baselines (180--2200\,km).

\section{Discussion} \label{discus}

The concentration of maser emission into two
compact clumps is likely a result of inhomogeneity in the
distribution and excitation of OH in the inner 100\,pc 
of \Z35, and not to compact structure in the background continuum.   
Chapman \etal\ (1990)
observed a $\lambda=18$\,cm continuum source with a flux density
of $\sim 30$\,mJy with MERLIN.
The non--detection of the continuum on our shortest
VLBA baselines places a lower limit on the angular
size of a 30\,mJy Gaussian source of 170\,mas, or 
an upper limit on the continuum brightness temperature, $T_C < 10^5$\,K.   
This suggests that the 18\,cm nuclear continuum structure 
is smooth on the 
angular scale of the maser clump separation.  
It is also possible that
filamentary compact continuum sources
of $<1$\,mJy are being amplified by the northern and southern maser clumps.
However, the components of the northern 
clump show a clear
gradient of velocity with position, suggesting that they
sample a coherent, possibly rotating, region of molecular gas (Figure~1,
{\it inset}).  50\% of the total maser flux density
was resolved in our VLBA images (Figure~3), which may indicate
an extended halo component of maser emission.
Smoothly distributed molecular gas could act as a low--gain
amplifier of an extended continuum source.
Alternatively, there may exist a population of compact maser clumps,
each with a flux density below our detection limit.  

If the
velocity gradient of 11\,\kms\,pc\inv\ that
we observe over 5\,pc in the northern clump (Figure~1; {\it inset})
is due to rotation of a homogeneous, gravitationally
bound cloud, 
the implied mass density is
$\sim 7000\,\msun$\,pc$^{-3}$ 
(total enclosed mass $\gsim 5\times 10^5\,\msun$).
The gradient suggested
in the line wings of the southern clump could be produced by rotation
of a cloud of similar density.
A difficulty in interpreting gradients in
spectral images arises when
two or more
distinct components are close in angle (less than a beamwidth) and velocity
(less than a linewidth).
Blending of the individual components can simulate a smooth velocity
gradient in an extended feature.
The MERLIN observations of
\cite{mon92} had a beamwidth of 280\,mas and a
channel spacing of 14\,\kms;  
at this resolution, 
spectral blending between the
two distinct maser clumps appeared as a gradient.  True gradients can
only be indentified with adequate resolution.
With this caution in
mind, we note that the {\sf S}--shaped structure in the northern maser
clump (Figure~2a) could be produced
by four or more blended, unresolved features.
Even so, these features would still display a correlation
between velocity and position.

Baan, Haschick, \& Henkel (1989) noted that the apparent
luminosity of extragalactic OH masers
scales approximately as $L^2_{\rm FIR}$.
They interpret this correlation to imply that
pumping by nuclear FIR radiation produces a column density of 
inverted OH that is proportional to $L_{\rm FIR}$, which
then acts as a low--gain amplifier of the 18\,cm radio continuum,
itself proportional to $L_{\rm FIR}$ (\cite{nor88}).  The apparent
optical depth estimated for the OH maser in \Z35,
$|\tau_{\rm OH}| \sim 1.7$, was based
on the assumption that the maser completely covers the background continuum
(\cite{cha90}).
Our observation indicates that the true covering factor 
is significantly less than unity.
We place an upper limit on the brightness
temperature of the 18\,cm background continuum,
$T_C \lsim 10^5$\,K, and a
lower limit on the peak brightness temperature of
the maser, $T_{\rm OH} > 3\times 10^6$\,K.  Assuming the
maser is unsaturated, we derive a lower limit on the
apparent optical depth, $|\tau_{\rm OH}| > 3.4$.   

If we assume that the maser is an unsaturated amplifier of background
continuum emission,
that the excitation temperatures of the 1667 and 1665--MHz
transitions are equal, and an LTE optical depth ratio
$\tau_{1667}/\tau_{1665}=1.8$, then 
the ratio of 1667 to 1665--MHz line peaks in single--dish spectra
of 9:1 implies 
$|\tau_{1667}| \sim 5$ (\cite{hen90}).
However, the two transitions do not peak at exactly the same velocities,
and the resulting variation in line ratios suggests
$3 \lsim |\tau_{1667}| \lsim 5$ across the emission profile.
Randell {\it et al.} (1995) compared the observed main--line
and satellite--line ratios with models of extragalactic OH emission regions,
and found that the OH line ratios are sensitive to a number of properties
of the molecular gas, including density, dust temperature and the
presence of velocity gradients.  The observed range of 1667 to 1665--MHz
line ratios may be due to
variations of these quantities within the maser clumps.
Our lower limit on the peak apparent optical depth, $|\tau_{1667}| > 3.4$,
is consistent with the
range of optical depths implied by the mainline flux ratios.
It therefore appears that low--gain models of extragalactic OH masers
({\it e.g.} \cite{baa89}) do not apply in \Z35.

One possible explanation for the angular structure of this OH maser is
that the clumps trace edge--on segments of molecular shocks.
Shocks are expected to be associated with energetic 
molecular outflows, such as
those inferred from the blueshifted line wings in several OH maser galaxies,
including \Z35 (\cite{baa89}).  The positions of OH masers in several Galactic
star--forming regions, 
{\it e.g.}, W3(OH) (\cite{blo92}), W75N (\cite{baa86}) and
G5.89--0.39 (\cite{woo93}), have been observed to coincide with shock fronts
in molecular outflows.   
The abundance of OH can be 
significantly enhanced in a post--shock
medium (see, \eg, \cite{eli78}; \cite{hol89}).  However, unlike
most extragalactic OH masers,
those in Galactic star--forming regions typically have 1667 to 1665--MHz
line ratios less than one (\cite{cas80}).  Mirabel, Rodr{\'\i}guez, \&
Ruiz (1989) have observed extended (pc--scale), 
weak OH maser emission in a number
of star--forming regions that exhibits line ratios closer to those seen
in extragalactic masers, and suggest that the two phenomena may be
related; if so, then a shock origin for the observed maser structure
would appear less likely.  
Detailed modelling (\cite{ran95}) also 
suggests that the brightest extragalactic OH masers
may arise in pc--scale
clouds with H$_2$ densities on the order of $10^4$\,cm$^{-3}$.
We note that images of OH absorption toward the center of our Galaxy
(\cite{boy94}) reveal a number
of ``wideline clouds'' that exhibit
velocity widths in excess of
100\,\kms, and intrinsic velocity gradients.  However, the typical size
of these clouds is rather larger ($\sim 100$\,pc),
and their velocity gradients at least an order of magnitude
smaller, than in the maser clumps in \Z35.

Although the continuous
structure inferred by Montgomery \& Cohen (1992) was probably a result of
spectral blending, it is still possible
that the two maser clumps lie in a rotating disk.  
We cannot rule out the presence of emission in the region between
the two clumps, given that 50\% of the total maser flux density was
resolved. 
The velocity separation of the maser peaks could be indicative of
rotation about a central mass of $2\times 10^7\,\msun$.
A difficulty with this model is that the
line profiles of the two clumps overlap in velocity by more than 50\,\kms, 
which would imply superimposed non--rotational velocities comparable
to the disk rotational velocity.
Furthermore, 
the velocity gradients we observe are more than an order of magnitude
greater than what we would expect from differential rotation alone.
More elaborate models, 
such as a warped disk, or emission on the surface of a thick torus,
cannot be ruled out by this observation, but are difficult to
quantify.
It is possible, despite the similar angular and velocity structure of
the clumps, that they are physically unconnected structures,
\eg, counter--rotating, gravitationally bound molecular
clouds, or shocks associated with a molecular outflow.

\section{Conclusion}

We have used the VLBA to image the OH maser in the nucleus of \Z35.
We detect maser emission in two clumps, with a projected
separation of 50\,pc along a position angle of $-10\deg$.  The northern
clump is extended, and exhibits a west--to--east
projected velocity gradient of 
$\sim 11$\,\kms\,pc\inv.  The brightest maser
emission arises in the southern clump, which is mostly compact ($<3$\,pc);
however, faint emission from the line wings suggests a velocity gradient
of similar magnitude to that in the northern clump,
but in the opposite direction.  
We detected no 18\,cm continuum emission in the nucleus,
which places a lower limit on the apparent
maser optical depth, $|\tau_{\rm OH}| > 3.4$, assuming the maser is
unsaturated and is amplifying continuum radiation.
This is the highest apparent gain yet 
determined for an extragalactic OH maser.  The observed flux ratios of 
the 1667 to 1665--MHz OH maser lines are consistent with optical depths
of this magnitude or greater.  Variations of the line ratio with angular
position and velocity suggest that physical conditions 
are not uniform within the maser clumps.
We did not find evidence for a 100--pc rotating, edge--on molecular
structure, as
reported by Montgomery \& Cohen (1992); however, since
50\% of the total maser flux was resolved, 
there is likely an extended or clumpy
component of maser emission in \Z35 that does not appear in our images.
Velocity gradients within the
clumps may indicate rotation on 
smaller scales (3--10\,pc).  The observed gradients
could be produced by two
counter--rotating, gravitationally bound molecular
clouds, each with an average density of
roughly $7000\,\msun$\,pc$^{-3}$.  Alternatively, the maser clumps
could be segments of molecular shocks viewed edge--on.

\clearpage

\figcaption[fg-zwall.eps]
{Distribution of 1667--MHz OH maser emission in the nucleus of IIIZw35.
The angular position offsets are referenced to the maser peak. 
Features greater than 5$\sigma$ (10\,mJy) in the
spectral images are indicated by circles; the diameter of each circle
is proportional to the flux density of the feature.  
The heliocentric velocity of the spectral peak in each clump is
indicated (optical definition;
$v({\rm Heliocentric, Optical}) = 
v({\rm LSR, Radio}) + 235$\,\kms\ at $v_{\rm sys}$).  
The directions
and ranges of the velocity gradients observed in each clump are indicated
with arrows.
The synthesized beam is $25\times 12$\,mas with a position angle
of $-27\deg$.  
{\it Inset}: A plot of velocity vs. R.A. offset
in the northern maser clump.
The solid line corresponds to the best--fit velocity gradient (over 
the range $8290<v<8340$\,\kms) of 
$5.6\pm 0.4$\,\kms\,mas\inv, or $11\pm 1$\,\kms\,pc\inv.}   

\figcaption[fg-zwexpand.eps]
{Expanded views of the OH maser clumps: 
(a) The northern clump; 
(b) The southern clump.
The diameter of each circle is proportional to the flux density of
the feature.
Selected features are labelled with their heliocentric velocities.  The
filled squares indicate the positions of the 1665--MHz emission peaks; their
heliocentric velocities are labelled in italics.}

\figcaption[fg-zwspec.eps]
{The 1667--MHz OH maser spectrum of IIIZw35.  The heavy solid curve
is a single--dish spectrum taken from Staveley--Smith {\it et al.} (1987).
The two solid curves show
the total imaged power as a function of heliocentric velocity
in each maser clump;
the dotted line is the sum of both clumps.  Arrows mark the velocities
of the 1665--MHz emission peaks. The marker at 8289\,\kms\ indicates
the systemic velocity.  Note that maser emission in single--dish spectra extends
to velocities beyond those plotted here.}


\begin{thebibliography}{}

\bibitem[Baan, Haschick, \& Henkel 1989]{baa89}
Baan, W. A., Haschick, A. D., \& Henkel, C. 1989, ApJ, 346, 680

\bibitem[Baart \etal\ 1986]{baa86}
Baart, E. E., Cohen, R. J., Davies, R. D., Rowland, P. R., \& Norris, R. P.
1986, MNRAS, 219, 145

\bibitem[Bloemhof, Reid, \& Moran 1992]{blo92}
Bloemhof, E. E., Reid, M. J., \& Moran, J. M. 1992, ApJ, 397, 500

\bibitem[Boyce \& Cohen 1994]{boy94}
Boyce, P. J. \& Cohen, R. J. 1994, A\&AS, 107, 563

\bibitem[Caswell, Haynes, \& Goss 1980]{cas80}
Caswell, J. L., Haynes, R. F., \& Goss, W. M. 1980, Australian J. Phys.,
33, 639

\bibitem[Chapman \etal\ 1990]{cha90}
Chapman, J. M., Staveley--Smith, L., Axon, D. J., Unger, S. W., 
Cohen, R. J., Pedlar, A., \& Davies, R. D. 1990, MNRAS, 244, 281

\bibitem[Elitzur \& De Jong 1978]{eli78}
Elitzur, M. \& De Jong, T. 1978, A\&A, 67, 323

\bibitem[Henkel \& Wilson 1990]{hen90}
Henkel, C. \& Wilson, T. L. 1990, A\&A, 229, 431

\bibitem[Hollenbach \& McKee 1989]{hol89}
Hollenbach, D. J. \& McKee, C. F. 1989, ApJ, 342, 306

\bibitem[Mirabel \& Sanders 1987]{mir87}
Mirabel, I. F. \& Sanders, D. B. 1987, ApJ, 322, 688

\bibitem[Mirabel, Rodr{\'\i}guez, \& Ruiz 1989]{mir89}
Mirabel, I. F., Rodr{\'\i}guez, L. F., \& Ruiz, A. 1989, ApJ, 346, 180

\bibitem[Montgomery \& Cohen 1992]{mon92}
Montgomery, A. S. \& Cohen, R. J. 1992, MNRAS, 254, 23P

\bibitem[Norris, Allen, \& Roche 1988]{nor88}
Norris, R. P., Allen, D. A., \& Roche, P. F. 1988, MNRAS, 234, 773

\bibitem[Randell \etal\ 1995]{ran95}
Randell, J., Field, D., Jones, K. N., Yates, J. A., \& Gray, M. D. 1995,
A\&A, 300, 659

\bibitem[Staveley--Smith \etal\ 1987]{sta87}
Staveley--Smith, L., Cohen, R. J., Chapman, J. M., Pointon, L.,
\& Unger, S. W. 1987, MNRAS, 226, 689

\bibitem[Wood 1993]{woo93}
Wood, D. O. S. 1993, in Astrophysical Masers, ed. A. W. Clegg \&
G. E. Nedoluha (Berlin: Springer--Verlag), 141

\bibitem[Zwicky 1971]{zwi71}
Zwicky, F., 1971, Catalogue of Selected Compact and Post Eruptive
Galaxies (Zurich: Speich)

\end{thebibliography}
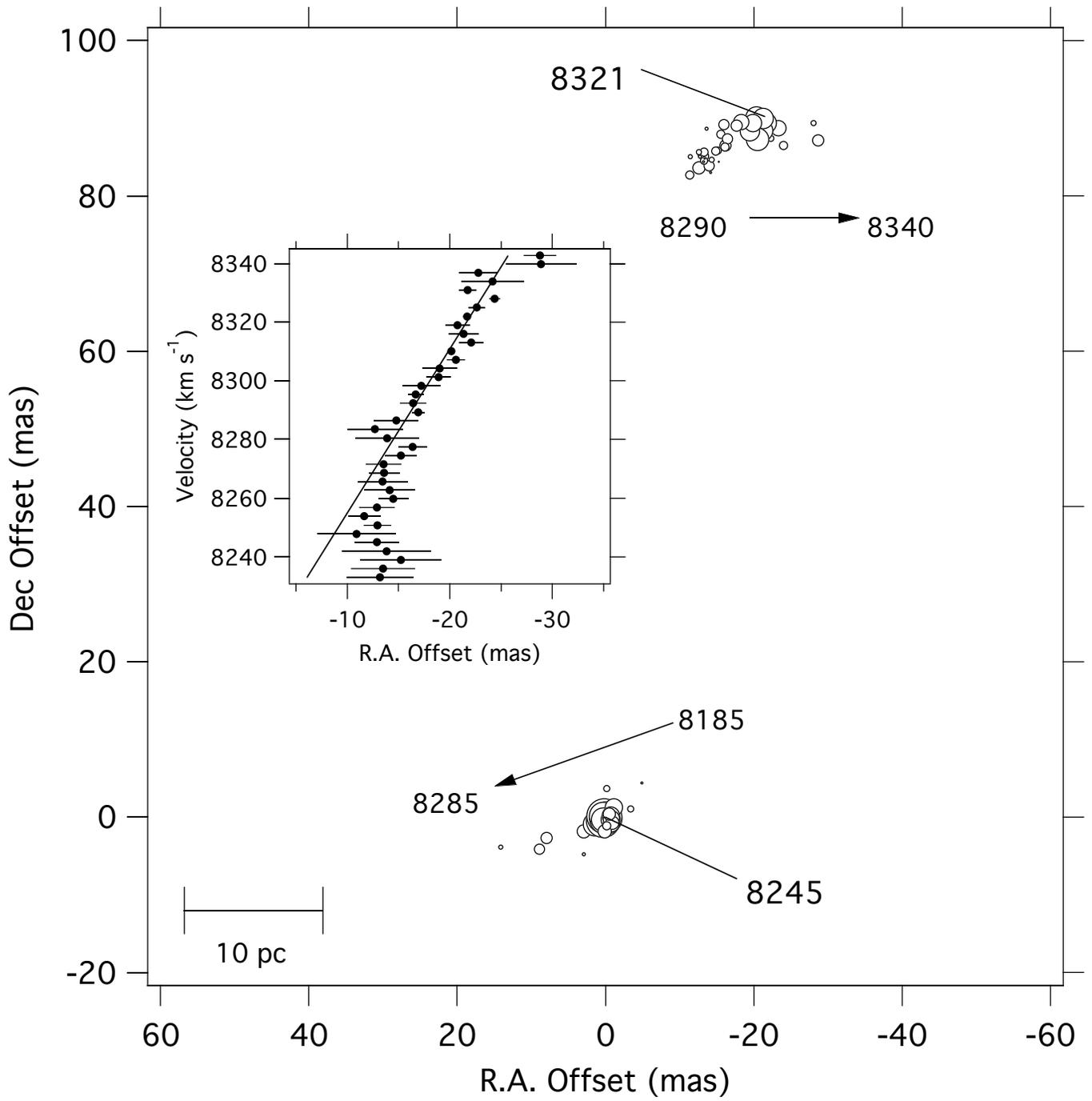
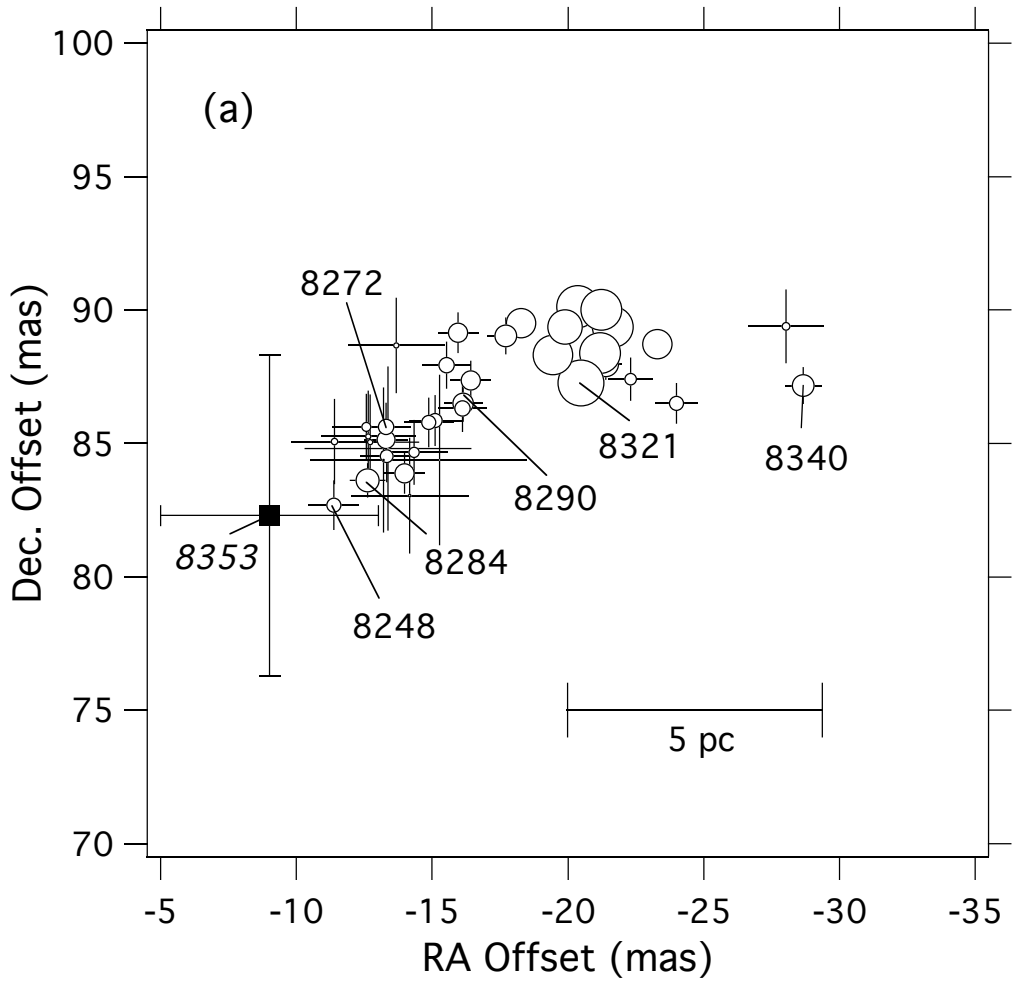
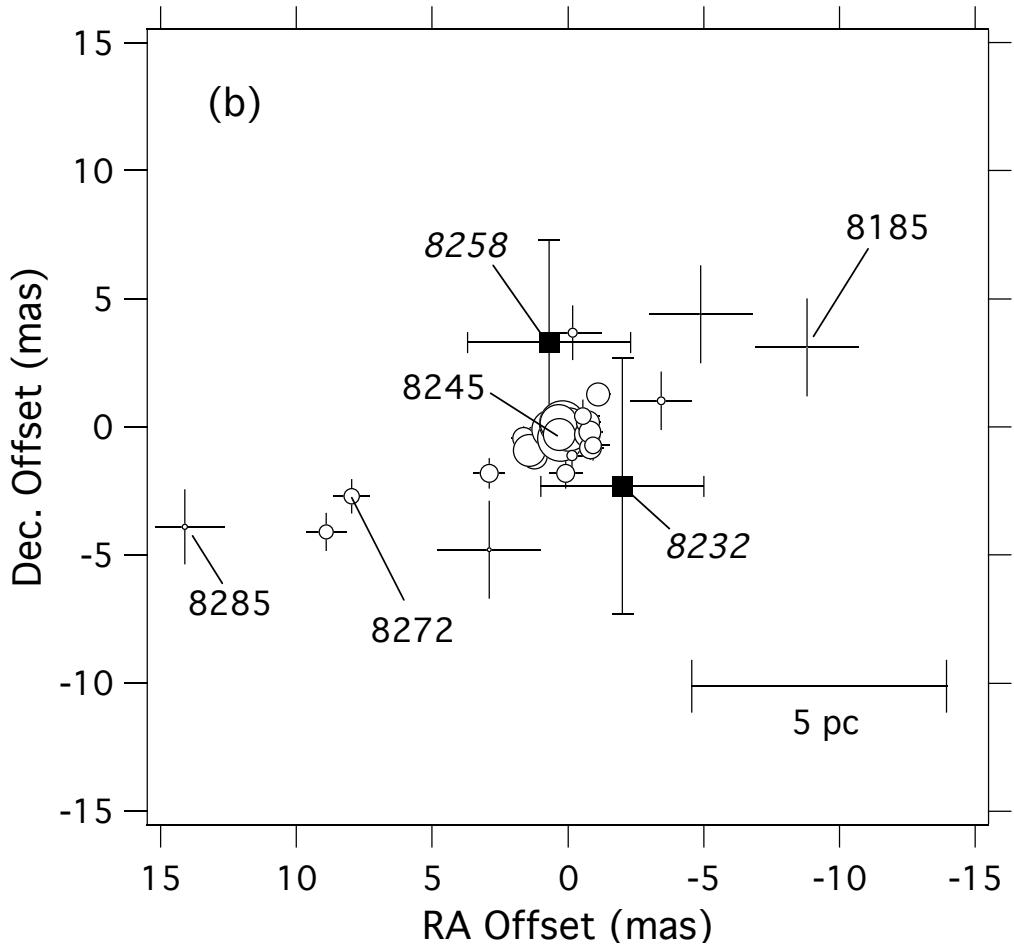
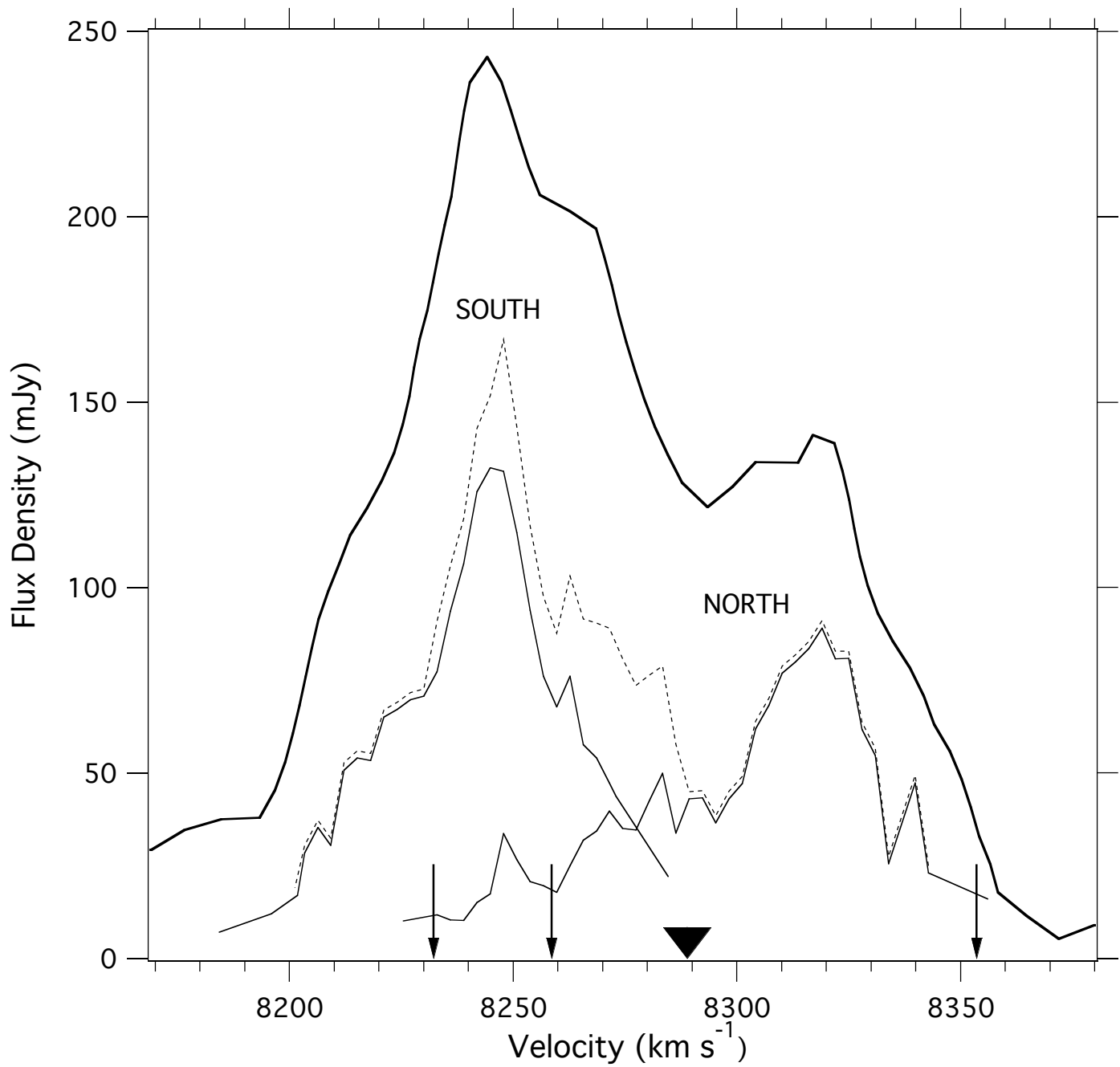
\end{document}